\documentclass[useAMS,usenatbib,usegraphicx]{mn2e}

\usepackage{times}
\usepackage{aas_macros}
\usepackage{amsfonts,amssymb}

\title[Dynamical State of Dark Matter Haloes I]
      {The Dynamical State of Dark Matter Haloes in Cosmological Simulations I:
      Correlations with Mass Assembly History}
      \author[Power, Knebe \& Knollmann]
      {Chris Power$^{1,2}$, Alexander Knebe$^{3}$ \& Steffen~R. Knollmann$^{3}$\\
        $^1$International Centre for Radio Astronomy Research, 
        University of Western Australia, 35 Stirling Highway, Crawley, 
        Western Australia 6009, Australia\\
        $^2$Department of Physics \& Astronomy, University of Leicester, 
        University Road, Leicester LE1 7RH, United Kingdom\\
        $^3$Departamento de F\'isica Te\'orica, M\'odulo C-15,
         Facultad de Ciencias, Universidad Aut\'onoma de Madrid,
         28049 Cantoblanco, Madrid, Spain}
\begin{document}
  
  \date{}
  
  \pagerange{\pageref{firstpage}--\pageref{lastpage}} \pubyear{}
  
  \maketitle
  
  \label{firstpage}
  
  \begin{abstract}

    Using a statistical sample of dark matter haloes drawn from a suite
    of cosmological $N$-body simulations of the Cold Dark Matter (CDM) model, 
    we quantify the impact of a simulated halo's mass accretion and merging 
    history on two commonly used measures of its dynamical state, the virial 
    ratio $\eta$ and the centre of mass offset $\Delta r$. Quantifying this
    relationship is important because the degree to which a halo is 
    dynamically equilibrated will influence the reliability with which we can 
    measure characteristic equilibrium properties of the structure and 
    kinematics of a population of haloes. We begin by verifying that a halo's 
    formation redshift $z_{\rm form}$ correlates with its virial mass 
    $M_{\rm vir}$ and we show that the fraction of its recently accreted mass 
    and the likelihood of it having experienced a recent major merger 
    increases with increasing $M_{\rm vir}$ and decreasing $z_{\rm form}$. We 
    then show that both $\eta$ and $\Delta r$ increase with increasing 
    $M_{\rm vir}$ and decreasing $z_{\rm form}$, which implies that 
    massive recently formed haloes are more likely to be dynamically unrelaxed 
    than their less massive and older counterparts. Our analysis shows that 
    both $\eta$ and $\Delta r$ are good indicators of a halo's dynamical 
    state, showing strong positive correlations with recent mass accretion 
    and merging activity, but we argue that $\Delta r$ provides a more
    robust and better defined measure of dynamical state for use in 
    cosmological $N$-body simulations at $z \simeq 0$. We find that 
    $\Delta r \la 0.04$ is sufficient to pick out dynamically relaxed haloes 
    at $z$=0. Finally, we assess our results in the context of previous 
    studies, and consider their observational implications.

  \end{abstract}
  
  
 \section{Introduction}

 One of the fundamental assumptions underpinning modern theories of galaxy 
 formation is that galaxies form and evolve in massive virialised haloes of 
 dark matter \citep[][]{white78,white91}. Characterising the properties of these
 haloes is an important problem, both theoretically and observationally, and its
 study has been one of the main objectives of cosmological $N$-body simulations
 over the last two decades. The majority of these simulations have modeled
 halo formation and evolution in a purely Cold Dark Matter (CDM) universe
 \citep[cf.][]{springel06}, with the focus primarily on their equilibrium 
 structure \citep[cf.][]{diemand09}. Various studies have revealed that CDM 
 haloes in dynamical equilibrium are triaxial structures 
 \citep[e.g.][]{bailin05} supported by velocity dispersion rather than rotation
 \citep[e.g.][]{bett07}, with mass profiles that are divergent down to the 
 smallest resolvable radius \citep[e.g.][]{diemand08,stadel09,navarro10} and 
 an abundance of substructure \citep[e.g.][]{diemand07,springel08b,gao11}.
 
 The qualification that a halo is in dynamical equilibrium is a particularly 
 important one when seeking to characterise the structure and kinematics of 
 simulated haloes in cosmological simulations. Previous studies have shown 
 that dynamically unrelaxed haloes tend to have lower central densities 
 \citep[see, for example,][]{tormen97,maccio07,romanodiaz07} and higher 
 velocity dispersions \citep[see, for example,][]{tormen97,hb06,donghia07} 
 than their dynamically relaxed counterparts. This means that a dynamically 
 unrelaxed halo is likely to have a measurably lower concentration 
 $c_{\rm vir}$ and higher spin parameter $\lambda$ than its dynamically relaxed 
 counterpart \citep[see, for example,][]{gardner01,maccio07}, 
 and so care must be taken to avoid contaminating halo samples with dynamically
 unrelaxed systems when measuring, for example, spin distributions 
 \citep[e.g.][]{bett07,maccio07,donghia07,Knebe08} and the correlation of 
 halo mass and concentration $c_{\rm vir}-M_{\rm vir}$ 
 \citep[e.g.][]{maccio07,neto07,gao07,Prada11}. 

 Yet haloes do not exist in isolation, and the degree to which they are 
 dynamically relaxed or unrelaxed bears the imprint of both their environment 
 and their recent mass assembly and merging history. As previous studies have 
 shown, dynamically unrelaxed haloes tend to have suffered one or more recent 
 significant mergers \citep[e.g.][]{tormen97,hb06}. For this reason, it is 
 common practice to use dynamical state and recent merging history 
 interchangeably, with the understanding implicit that unrelaxed haloes 
 are ones that have suffered one or more recent major mergers.

 However, it is important to establish this practice on a more quantitative 
 footing and to assess how well a halo's dynamical state and its 
 recent mass assembly history correlate. This is because of the need to 
 identify robustly haloes that are in dynamical equilibrium -- or indeed 
 disequilibrium -- in cosmological simulations\footnote{Our focus is fixed 
   firmly on haloes in cosmological simulations, but we note that the 
   relationship between dynamical
   state and recent mass assembly history is equally important observationally.
   Here, for example, estimates of the dynamical masses of galaxy clusters 
   require assume a population of dynamical tracers that are in dynamical 
   equilibrium \citep[e.g.][]{piffaretti08}, while reconstructions of a 
   galaxy cluster's recent merging history look for signatures of 
   disequilibrium \citep[e.g.][]{cassano10}. See \S~\ref{sec:summary} for 
   further discussion.}. The goal of this paper is to quantify this
 relationship using a statistical sample of haloes drawn from cosmological
 $N$-body simulations of the CDM model. The CDM model is the ideal testbed 
 for this study because of the fundamental role merging plays in halo mass
 assembly 
 \citep[e.g.][]{maulbetsch07,fakhouri08,mcbride09,fakhouri10,fakhouri10a}, and 
 because we expect  massive haloes, which on average form later than their 
 less massive counterparts, to have more violent recent merging histories. 

 Such an undertaking has practical implications. For example, if we want to 
 robustly characterise the predicted variation of, say, concentration 
 $c_{\rm vir}$ with virial mass $M_{\rm vir}$ on galaxy group and cluster mass 
 scales ($M_{\rm vir} \ga 10^{13} \rm M_{\sun}$), then it is essential that we 
 can identify relaxed systems in a robust fashion. Should we use mass assembly 
 histories directly and select only haloes that have quiescent recent merging 
 histories, or are commonly used measures that estimate dynamical state based 
 on material within the halo's virial radius $r_{\rm vir}$ adequate? This is 
 particularly important for comparison with observations that provide crucial 
 tests of the theory, such as the analysis the $M_{\rm vir} - c_{\rm vir}$ 
 relation for groups and clusters drawn from the Sloan Digital Sky Survey by 
 \citet{mandelbaum08}.

 In this paper, we examine how a halo's mass assembly history and dynamical 
 state varies with its virial mass $M_{\rm vir}$ and its formation redshift, and
 adopt simple measures to characterise a halo's recent mass assembly and 
 merging history -- namely, the fraction of mass assembled ($\Delta M/M$); the 
 rate of change of mass with redshift $1/M dM/dz$; and the most significant 
 merger $\delta_{\rm max}$. We compare these with two measures of the halo's 
 dynamical state -- the virial ratio
 \begin{equation}
   \eta=2T/|W| ,
 \end{equation}
 \noindent where $T$ and $W$ are the kinetic and gravitational 
 potential energies of halo material \citep[cf.][]{colelacey96,hb06},
 and the centre-of-mass offset
 \begin{equation}
   \Delta r=|\vec{r}_{\rm cen}-\vec{r}_{\rm cm}|/r_{\rm vir}, 
 \end{equation}
 \noindent where $\vec{r}_{\rm cen}$ and $\vec{r}_{\rm cm}$ are the 
 centres of density and mass of halo material and $r_{\rm vir}$ is the 
 halo's virial radius \citep[cf.][]{crone96,thomas98,thomas01}. 
 Previous studies have shown that both $\eta$ and $\Delta r$ 
increase in the aftermath of a major merger  \citep[e.g.][]{hb06,poole06}, and 
 we will clarify precisely how they relate to a halo's mass assembly and 
 merging activity in general. We note that our work develops earlier ideas
 presented in \citet{Knebe08}, in which we investigated the relationship 
 between halo mass $M_{\rm vir}$ and spin $\lambda$, and it complements that of 
 \citet{Davis11}, who address related but distinct issues in their critique of 
 the application of the virial theorem (cf. \S\ref{ssec:virial_theorem}) to 
 simulated high redshift dark matter haloes.\\

 The layout of the paper is as follows. In \S\ref{sec:sims}, we
 describe our approach to making initial conditions; finding and
 analysing dark matter haloes in evolved outputs; constructing
 merger trees of our dark matter haloes; and our criteria for defining our
 halo sample. In \S\ref{sec:hierarchical}, we examine the relationship 
 between a halo's virial mass $M_{\rm vir}$, its formation time $z_{\rm form}$ 
 and measures of its mass accretion and merging history. In 
 \S\ref{sec:equilibrium}, we present commonly used measures for assessing
 the dynamical state of a dark matter halo -- the virial ratio $\eta=2T/|W|$
 (cf. \S\ref{ssec:virial_theorem}) and the centre-of-mass offset 
 $\Delta r= |\vec{r}_{\rm cen}-\vec{r}_{cm}|/r_{\rm vir}$ (cf. 
 \S\ref{ssec:centre_of_mass_offset}) -- and investigate how these measures
 correlate with $M_{\rm vir}$ and $z_{\rm form}$. In 
 \S\ref{sec:linking_mass_assembly_and_dynamical_state} we combine the insights
 from the previous two sections and show how a halo's dynamical state depends
 on its recent mass accretion and merging history. Finally, we summarise our 
 results in \S\ref{sec:summary} and comment on the implications of our findings
 for both observational studies and galaxy formation modeling.

\section{Methods}
\label{sec:sims}

\subsection{The Simulations}
\label{ssec:halo_ics}

We have run a series of $256^3$ particle cosmological $N$-body simulations 
following the formation and evolution of structure in the CDM model. We use
a sequence of boxes of side $L_{\rm box}$ varying between $20 h^{-1} \rm Mpc$ 
and $500 h^{-1} \rm Mpc$ from $z_{\rm start}$=100 to $z_{\rm finish}$=0. In
each case we assume a flat cosmology with a dark energy term, with
cosmological parameters $\Omega_0=0.7$, $\Omega_{\Lambda}=0.3$, $h=0.7$, and 
a normalisation $\sigma_8=0.9$ at $z=0$. Various properties of these 
simulations are summarised in table~\ref{tab:sim_props}. 

Initial conditions were generated using a standard procedure that can be
summarised as follows;

\begin{enumerate}
\item Generate the CDM transfer function for the appropriate cosmological 
  parameters ($\Omega_0,\Omega_{\Lambda}, \Omega_{\rm b}$ and $h$) using the 
  Boltzmann code {\small CMBFAST} \citep[][]{cmbfast}. This is convolved with 
  the primordial power spectrum $P(k) \propto k^n$, $n=1$, to obtain the
  unnormalised power spectrum, which is normalised by requiring that the 
  linear mass $\sigma(R)$ equal $\sigma_8$ on a scale of $8 h^{-1} \rm Mpc$ at 
  $z$=0.
\item Create a statistical realisation of a Gaussian random field of
  density perturbations in Fourier space, whose variance is given by
  $P(k)$, where $k=\sqrt{k_x^2+k_y^2+k_z^2}$ and whose mean is zero. 
\item Take the inverse transform of the density field and compute
  positions and velocities using the Zel'dovich approximation.
\item Impose these positions and velocities on an initial uniform particle
  distribution such as a grid or ``glass''.
\end{enumerate}

\noindent Note that throughout our we use a ``glass''-like
configuration as our initial uniform particle distribution
\citep{White96}.\\

All simulations were run using the parallel TreePM code {\small GADGET2} 
\citep{gadget2} with constant comoving gravitational softening $\epsilon$ and
individual and adaptive timesteps for each particle, 
$\Delta t = \eta \sqrt{\epsilon/a}$, where $a$ is the magnitude of a 
particle's gravitational acceleration and $\eta=0.05$ determines the accuracy 
of the time integration.

\begin{table}
  \begin{center}
    \caption{\textbf{Properties of the Simulations}. Each of the simulations 
      contains $256^3$ particles. In addition, $L_{\rm box}$ is the comoving 
      box length in units of $h^{-1} \rm Mpc$; $\rm N_{\rm run}$ is the
      number of runs in the series; $m_{\rm part}$ is the particle mass in 
      units of $h^{-1} {\rm M}_{\sun}$; $\epsilon$ is the force softening 
      in comoving units of $h^{-1} \rm kpc$; and $M_{\rm cut}$ is the halo mass
      corresponding to $N_{\rm cut}$=600 particles, in units of $h^{-1} 
      {\rm M}_{\sun}$.}
    
    \vspace*{0.3 cm}

    \begin{tabular}{lccccc}
      \hline
      Run & $L_{\rm box}$ & ${\rm N}_{\rm run}$ & $m_{\rm part}$
      & $\epsilon$ & $M_{\rm cut}$\\
      \hline 
      
      $\Lambda$CDM\_L20& $20$  & 5 & $3.97\times10^7$ & 1.5 & $2.38 \times 10^{10}$\\
      
      $\Lambda$CDM\_L50& $50$  & 1 & $6.20\times10^8$ & 3.9 & $3.72 \times 10^{11}$\\
      
      $\Lambda$CDM\_L70& $70$  & 1 & $1.70\times10^9$ & 5.5 & $10^{12}$\\
      
      $\Lambda$CDM\_L100& $100$ & 1 & $4.96\times10^9$ & 7.8 & $2.97 \times 10^{12}$\\
      
      $\Lambda$CDM\_L200& $200$ & 1 & $3.97\times10^{10}$ & 15.6 & $2.39 \times 10^{13}$\\
      
      $\Lambda$CDM\_L500& $500$ & 1 & $6.20\times10^{11}$ & 39.1 & $3.72 \times 10^{14}$\\
      \hline
    \end{tabular}
    \label{tab:sim_props}
  \end{center}
\end{table}

\subsection{Halo Identification and Merger Trees}
\label{ssec:halo_properties}

\paragraph*{Halo Catalogues}
Groups were identified using the MPI-enabled version of {\small AHF}, 
otherwise known as {\small{\textbf{A}MIGA}'s \small{\textbf{H}}alo 
\small{\textbf{F}}inder}\footnote{{\small AHF} may be downloaded from 
  \small{\texttt{http://popia.ft.uam.es/AMIGA}}} \citep{Knollmann09}. {\small AHF} is a 
modification of {\small MHF} \citep[{\small MLAPM}'s {\small H}alo 
  {\small F}inder; see][]{gill04} that locates groups as peaks
in an adaptively smoothed density field using a hierarchy of grids and a 
refinement criterion that is comparable to the force resolution of the 
simulation. Local potential minima are calculated for each of these
peaks and the set of particles that are gravitationally bound to the
peaks are identified as groups that form our halo catalogue.

For each halo in the catalogue we determine its centre-of-density 
$\vec{r}_{\rm cen}$ (using the iterative ``shrinking spheres'' method described 
in \citealt{power03}) and identify this as the halo centre. From this, we 
calculate the halo's virial radius $r_{\rm vir}$, which we define as the radius 
at which the mean interior density is $\Delta_{\rm vir}$ times the critical 
density of the Universe at that redshift, $\rho_{\rm c}(z)=3H^2(z)/8\pi G$, 
where $H(z)$ and $G$ are the Hubble parameter at $z$ and the gravitational 
constant respectively. The corresponding virial mass $M_{\rm vir}$ is 
\begin{equation}
  \label{eq:mvir}
        {M_{\rm vir}=\frac{4\pi}{3} \Delta_{\rm vir} \rho_{\rm c} r_{\rm vir}^3.}
\end{equation}

\noindent We adopt a cosmology- and redshift-dependent overdensity criterion, 
which for a $\Lambda$CDM cosmology with $\Omega_0=0.3$ and 
$\Omega_{\Lambda}=0.7$ gives $\Delta_{\rm vir}\simeq 97$ at $z$=0
\citep[c.f.][]{eke98}. 

\paragraph*{Merger Trees}
Halo merger trees are constructed by linking halo particles at consecutive
output times;

\begin{itemize}
\item For each pair of group catalogues constructed at consecutive
  output times $t_1$ and $t_2>t_1$, the ``ancestors'' of ``descendent''
  groups are identified. For each descendent identified in the catalogue at
  the later time $t_2$, we sweep over its associated particles and
  locate every ancestor at the earlier time $t_1$ that contains a subset
  of these particles. A record of all ancestors at $t_1$ that contain
  particles associated with the descendent at $t_2$ is maintained.

\item The ancestor at time $t_1$ that contains in excess of $f_{\rm prog}$
  of these particles and also contains the most bound particle of the
  descendent at $t_2$ is deemed the \emph{main progenitor}. Typically $f_{\rm
  prog}=0.5$, i.e. the main progenitor contains in excess of half the
  final mass.

\end{itemize}

Each group is then treated as a node in a tree structure, which can be
traversed either forwards, allowing one to identify a halo at some
early time and follow it forward through the merging hierarchy, or
backwards, allowing one to identify a halo and all its progenitors at
earlier times. 

\subsection{Defining the Halo Sample}
\label{ssec:halo_sample}

A degree of care must be taken when choosing which haloes to include in our
sample, to ensure that our results are not affected by the finite resolution
of our simulations. One of the key calculations in this study is of a halo's 
virial ratio $\eta=2T/|W|$ (see \S~\ref{ssec:virial_theorem}), where $T$ and 
$W$ are the kinetic and gravitational potential energies of material within 
$r_{\rm vir}$. The gravitational potential energy is particularly sensitive to 
resolution; if a halo is resolved with too few particles, its internal 
structure will not be recovered sufficiently accurately and the magnitude of 
$W$ will be underestimated. 

\begin{figure}
\includegraphics[width=8cm]{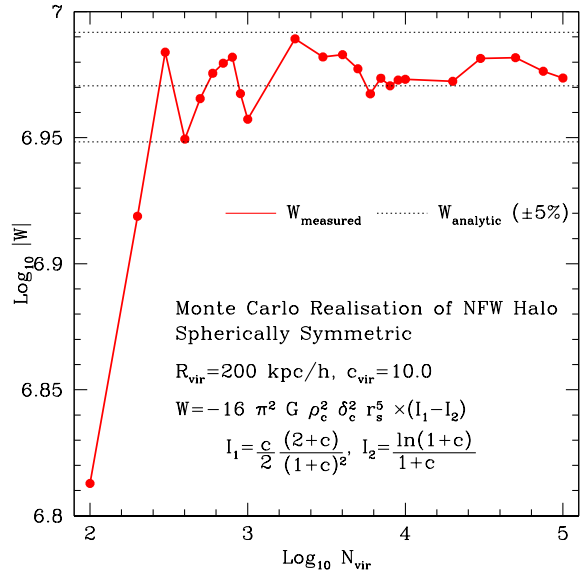}
\caption{{\bf How many particles are required to measure accurately the 
    gravitational binding energy of a dark matter halo?}. Here we generate
Monte Carlo realisations of a NFW halo and calculate the gravitational 
potential energy of material within the virial radius. If there are too few 
particles within $r_{\rm vir}$, the potential energy will be inaccurate.}
\label{fig:monte_carlo_energy} 
\end{figure}

We estimate how many particles are needed to recover $W$ robustly from a 
$N$-body simulation in Figure~\ref{fig:monte_carlo_energy}. Here we generate 
Monte Carlo $N$-body realisations of a halo whose spherically averaged mass 
profile is described by the \citet{nfw97} profile,
\begin{equation}
\frac{\rho(x)}{\rho_{\rm c}} = \frac{\delta_{\rm c}}{cx\,(1+cx)^2};
\end{equation}
\noindent here $x=r/r_{\rm vir}$ is the radius $r$ normalised to $r_{\rm vir}$,
$c$ is the concentration parameter and $\delta_{\rm c}$ is the characteristic
density,
\begin{equation}
\delta_{\rm c} = \frac{\Delta_{\rm vir}}{3}\frac{c^3}{\ln(1+c)-c/(1+c)}.
\end{equation}
\noindent The resulting gravitational potential energy is given by
\begin{equation}
\label{eq:wnfw}
W = -16\pi^2\,G\,\rho_{\rm c}^2\,\delta_c^2\,\left(\frac{r_{\rm vir}}{c}\right)^5 \times \left[\frac{c}{2}\frac{(2+c)}{(1+c)^2}-\frac{\ln(1+c)}{(1+c)}\right].
\end{equation}
\noindent In a $N$-body simulation or realisation, we calculate $W$ by randomly 
sampling particles within $r_{\rm vir}$ and rescaling; this gives
\begin{equation}
  \label{eq:potential}
  W=\left(\frac{N_{\rm vir}^2-N_{\rm vir}}{N_k^2-N_k}\right)\left(\frac{-Gm_{\rm p}^2}{\epsilon}\right)\Sigma_i^{N_k-1}\Sigma_{j=i+1}^{N_k} -K_s(|r_{ij}|/\epsilon),
\end{equation}
\noindent where there are $N_{\rm vir}$ particles in the halo, each of mass
$m_{\rm p}$. We sample $N_k$ particles from $N_{\rm vir}$, $|r_{ij}|$ 
is the magnitude of the separation between particles $i$ at $\vec{r_i}$ and 
$j$ at $\vec{r_j}$, and the prefactor $(N^2-N)/(N_k^2-N_k)$ accounts for 
particle sampling. $\epsilon$ is the gravitational softening and $K_s$ 
corresponds to the softening kernel used in {\small GADGET2}. For the Monte
Carlo realisations in Figure~\ref{fig:monte_carlo_energy} we set 
$\epsilon$ to be vanishingly small, but for the simulations we use $\epsilon$ 
as it is listed in table~\ref{tab:sim_props}.

Figure~\ref{fig:monte_carlo_energy} shows $|W|$ measured for Monte Carlo 
realisations of a halo with $c\!=\!$10 and $r_{\rm vir}$=200 kpc as a function of
$N_{\rm vir}$. For comparison the horizontal dotted lines indicate the value of 
$|W|$ ($\pm 5\%$) we expect from equation (\ref{eq:wnfw}). If 
$N_{\rm vir} \approx 300$ or fewer, the measured $|W|$ deviates from the 
expected $|W|$ by greater than $5\%$; therefore we might regard 
$N_{\rm cut}=300$ as the lower limit on $N_{\rm vir}$ for a halo to be
included in our sample. However, we adopt a more conservative $N_{\rm cut}=600$
in the remainder of this paper; this is because the structure of simulated 
haloes are affected by finite gravitational softening \citep[cf.][]{power03}, 
they are are seldom (if ever) smooth and spherically symmetric 
\citep[e.g.][]{bailin05}, as we assumed in our simple calculation, and there 
can be a range of concentrations at a given mass \citep{bullock01a}, which 
will affect any estimate of $W$ as inspection of equation (\ref{eq:wnfw})
reveals. 

\section{Quantifying Mass Assembly \& Merging History}
\label{sec:hierarchical}

In this section we establish quantitative measures for a halo's mass accretion 
and merging histories, and we examine how these measures relate to virial
mass $M_{\rm vir}$ and formation redshift $z_{\rm form}$.

\paragraph*{Quantifying Formation Redshift}

\begin{figure} 
  \includegraphics[width=8cm]{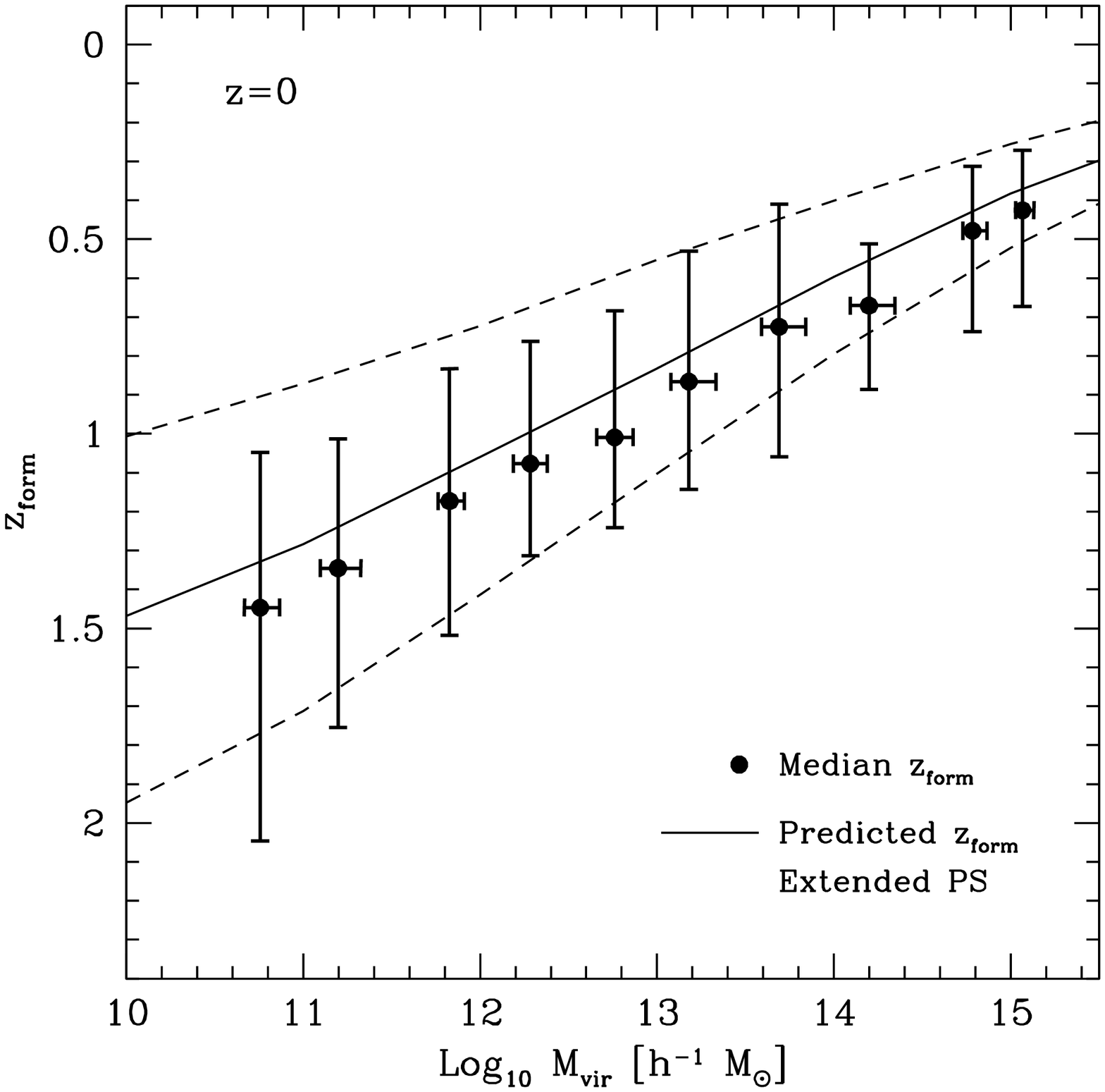}
\caption{\textbf{Relationship between Virial Mass and Formation Redshift}.
  Here we show how the formation redshift $z_{\rm form}$ varies
  with virial mass $M_{\rm vir}$ at $z$=0. We determine $z_{\rm form}$ directly
  from a halo's merger tree -- for a halo of $M_{\rm vir}$ identified
  at $z$=0, we identify the redshift $z_{\rm form}$ at which the mass
  of its main progenitor first exceeds half its virial mass at
  $z$=0. Data are binned using equally spaced bins in Log$_{10} M_{\rm vir}$;
  filled circle and bars correspond to medians, upper and lower
  quartiles. The solid, upper and lower dashed curves corresponds to
  the median $z_{\rm form}$ and its upper and lower quartiles predicted
  by extended Press-Schechter theory \citep[cf.][]{laceycole93}.}
\label{fig:mass_versus_zform}
\end{figure}

We begin our analysis by verifying the correlation between virial mass 
$M_{\rm vir}$ and formation redshift $z_{\rm form}$ for our halo sample.
We adopt the convention of \citet{colelacey96} and define $z_{\rm form}$ 
as the redshift at which the mass of the main progenitor of a halo of 
mass $M_{\rm vir}(z)$ identified at $z$ first exceeds $M_{\rm vir}(z)/2$. This
is equivalent to $z_{\rm 1/2,mb}$ in the survey of halo formation redshift 
definitions examined by \citet{li08}. 

Our expectation is that more massive CDM haloes will assemble more of 
their mass at later times than their lower mass counterparts and this is borne
out by Figure~\ref{fig:mass_versus_zform}. Here we show the variation of 
$z_{\rm form}$ with $M_{\rm vir}$ for our halo sample; the filled circles and 
bars indicate the medians and upper and lower quartiles respectively, within 
logarithmic mass bins of width 0.5 dex. The relationship between the mean
and median $z_{\rm form}$ with $M_{\rm vir}$ can be well approximated by 
\begin{equation}
  \left<z_{\rm form}\right> \simeq -0.22\,\log_{10} M_{12}+1.06,
\end{equation}
\noindent and 
\begin{equation}
  {\mathrm {Med}} z_{\rm form} \simeq -0.23\,\log_{10} M_{12}+1.08,
\end{equation}
where $M_{12}$ is $M_{\rm vir}$ in units of $10^{12} h^{-1}{\rm M}_{\sun}$.
This is in very good agreement with the mean variation reported 
for the ``Overall'' sample of haloes drawn from the Millennium and 
Millennium II simulations (cf. \citealt{springel.etal.2005} and 
\citealt[][]{boylan.etal.2009} respectively) in Table 3 of 
\citet{mcbride09}, who found
\[ \left<z_{\rm form}\right>=-0.24\,\log_{10} M_{12}+1.26 .\]
\noindent We show also the variation predicted by extended Press-Schechter 
(EPS) theory for our choice of CDM power spectrum -- see the solid and dashed 
curves, indicating the median, upper and lower quartiles of the distributions 
\citep[cf.][]{laceycole93}. These curves were generated using realisations 
of $10^6$ Monte-Carlo merger trees for haloes with $z$=0 masses in the range 
$10^{10} \leqslant M_{\rm vir}/h^{-1} M_{\sun} \leqslant 10^{15.5}$. We note a 
slight but systematic offset between the medians evaluated from the simulated 
haloes and those predicted by EPS theory, such that the simulated haloes tend 
to form earlier than predicted. This effect has been reported previously by both
\citet{vandenbosch02} and \citet{maulbetsch07}.
 
\paragraph*{Quantifying Recent Mass Accretion History}
Because more massive systems tend to form later than their less
massive counterparts, it follows that the rate at which a halo
assembles its mass should increase with increasing $M_{\rm vir}$
and decreasing $z_{\rm form}$. The recent comprehensive study
by \citet{mcbride09} provides a useful fitting formula that captures
the complexity of a halo's mass accretion history and allows haloes 
to be categorised into different Types I to IV, which depend on their 
growth rates. However, we adopt two simple well-defined measures of 
a halo's mass accretion rate that have a straightforward interpretation;
\begin{itemize}
\item $(\Delta M/M)_{\Delta t}$, the fraction of mass that has been accreted 
  by a halo during a time interval $\Delta t$; and
\item $\alpha=1/M dM/dz$, the rate of fractional change in a halo's virial 
  mass with respect to redshift over a redshift interval $\Delta z$.
\end{itemize}

\noindent Note that $\alpha$ is equivalent to the $\alpha$ free parameter 
used in \citet{wechsler02}. We find that $(\Delta M/M)_{\Delta t}$ and 
$\alpha$ are sufficient as simple measures of the mass accretion rate and 
we use them in the remainder of this paper.

For the fiducial timescale $\Delta t$, we use twice the dynamical timescale
$\tau_{\rm dyn}$ estimated at the virial radius, 
\begin{equation}
  \label{eq:tdyn}
  \tau_{\rm dyn} = \sqrt{2}\frac{r_{\rm vir}}{V_{\rm vir}}
  = 2.8\,\left(\frac{\Delta_{\rm
      vir}}{97}\right)^{-1/2}\left(\frac{H(z)}{70}\right)^{-1} \rm Gyrs
\end{equation}
\noindent Note that $\tau_{\rm dyn}$ depends only on $z$ and is the same for 
all haloes. For our adopted cosmological parameters, $\Delta(z) \simeq 97$ at 
$z$=0, and so $\Delta t = 2\,\tau_{\rm dyn} \simeq 5.6$ Gyr which corresponds 
to a redshift interval of $\Delta z \simeq 0.6$ at $z$=0. Merging proceeds
on a timescale $\tau_{\rm merge} \ga \tau_{\rm dyn}$, with 
$\tau_{\rm merge} \rightarrow \tau_{\rm dyn}$ as the mass ratio of the 
merger decreases. Our adopted timescale of $\Delta t=2\tau_{\rm dyn}$ for the 
response of a halo to a merger is reasonable when compared to typical values 
of $\tau_{\rm merge}/\tau_{\rm dyn}$ expected for haloes in cosmological 
simulations, as estimated by \citet{boylanetal08}\footnote{In 
  particular, we refer to their equation 5 with values of 
  $j/j_{\rm C}(E)=0.5$ and $r_{\rm C}(E)/r_{\rm vir}$ that are consistent 
  with the results of cosmological simulations. Here $j$ is the specific 
  angular momentum of a merging subhalo, $j_{\rm C}(E)$ is the specific 
  angular momentum of the circular orbit corresponding to the subhalo's 
  orbital energy $E$, and $r_{\rm C}(E)$ is the radius corresponding 
  to this circular orbit.}.

We determine both $(\Delta M/M)_{\tau_{\rm dyn}}$ and $\alpha$ directly from
each halo's merger tree by tracking $M_{\rm vir}(z)$ of its main progenitor 
over the interval $\Delta z$; $\alpha$ is obtained by taking the natural 
logarithm of the progenitor mass at each redshift and estimating its value 
by linear regression. Haloes that have high mass accretion rates will have 
$(\Delta M/M)_{\Delta t} \rightarrow 1$ and $\alpha \rightarrow -\infty$.

\begin{figure}
  \includegraphics[width=8cm]{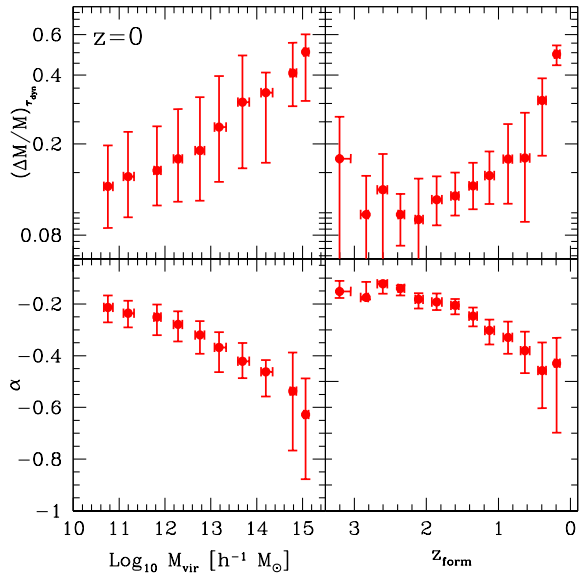}
  \caption{\textbf{Relationship between Recent Mass Accretion History,
      Halo Mass and Formation Redshift}. For each halo of virial mass 
    $M_{\rm vir}$ and formation redshift $z_{\rm form}$ identified at 
    $z$=0, we follow its merger tree back for one dynamical time 
    $\tau_{\rm dyn}$ ($\simeq 4.6$ Gyrs, $\Delta z \simeq 0.45$) and 
    characterise its mass accretion history using two measures. The
    first is $(\Delta M/M)_{\tau_{\rm dyn}}$, the fraction of mass 
    accreted over $\tau_{\rm dyn}$ (upper panels), and the second is 
    $\alpha$, the average mass accretion rate of \citet{wechsler02}
    (lower panels). Data points and bars correspond to medians
    and upper and lower quartiles. Note that we use
    equally-spaced logarithmic bins in $M_{\rm vir}$ and $z_{\rm form}$.}
\label{fig:mass_versus_accretion_measures}
\end{figure}

In Fig.~\ref{fig:mass_versus_accretion_measures} we show how a halo's mass 
accretion rate correlates with its virial mass and formation time. 
$(\Delta M/M)_{\tau_{\rm dyn}}$ ($\alpha$) shows a steady monotonic 
increase (decrease) as $M_{\rm vir}$ increases over the range 
$10^{10} h^{-1} \rm M_{\sun} \la M_{\rm vir} \la 10^{15} h^{-1} \rm 
M_{\sun}$. For example, inspection of $(\Delta M/M)_{\tau_{\rm dyn}}$ reveals that 
$\la 15\%$ of the virial mass of a halo with 
$M_{\rm vir} \sim 10^{12} h^{-1} M_{\sun}$ has been accreted since 
$z \simeq 0.5$, compared to $\sim 50\%$ for haloes with 
$M_{\rm vir} \sim 10^{15} h^{-1} M_{\sun}$ over the same period.  
$(\Delta M/M)_{\tau_{\rm dyn}}$ ($\alpha$) shows a similar increase (decrease) 
with decreasing $z_{\rm form}$ although it's interesting to note that the trend
flattens off for haloes that form at $z \ga 2$.

This analysis confirms our theoretical prejudice that more massive haloes 
and haloes that formed more recently tend to be the haloes with the 
measurably highest accretion rates. Reassuringly, our results are in 
good agreement with the findings of recent studies. For example, 
\citet{mcbride09} examined the mass accretion and merging histories 
of a much larger sample of haloes drawn from the Millennium and Millennium-II 
simulations and found that the mean instantaneous mass accretion rate varies 
with halo mass as $\dot{M}/M\propto\,M^{0.127}$; this compares favourably with 
our equivalent measure, $(\Delta M/M)_{\tau_{\rm dyn}}\propto\,M_{\rm vir}^{0.14}$.
\citet{maulbetsch07} looked at halo accretion rates, normalised to their
maximum masses, over the redshift interval $z$=0.1 to 0 for haloes with masses 
$10^{11} \leq M_{\rm vir}/{h^{-1} \rm M_{\sun}} \leq 10^{13}$ and found only a weak
dependence on halo mass, with higher mass haloes have higher rates. This is 
consistent with with our results for $\alpha$, whose median value changes by 
$\sim 10\%$ over the same range in halo mass.

\begin{figure}
  \includegraphics[width=8cm]{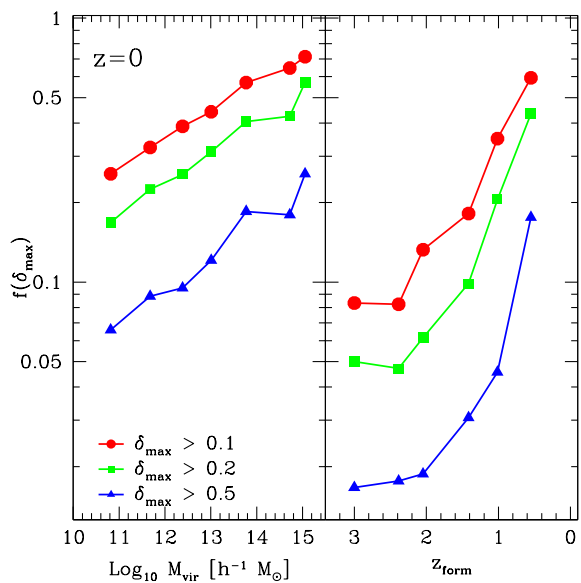}
  \caption{\textbf{Frequency of Major Mergers and Dependence on 
      Halo Mass and Formation Redshift}. Here we determine the
    most significant merger of mass ratio $\delta_{\rm max}=M_{\rm
      acc}(z_i)/M_{\rm vir}(z_f)$ experienced by each halo since
    $z$=0.5, where $z_i$ and $z_f$ correspond to the initial and
    final redshifts. We then compute the fraction of haloes 
    $f(\delta_{\rm max})$ at a given virial mass (left hand panel) 
    and given formation redshift (right hand panel) that have
    experienced mergers with mass ratios $\delta_{\rm max}$ in 
    excess of $10\%$ (filled circles), $20\%$ (filled squares) 
    and $50\%$ (filled triangles).}
\label{fig:frequency_of_mergers}
\end{figure}

\paragraph*{Quantifying Recent Merger Activity}
Both $(\Delta M/M)_{\tau_{\rm dyn}}$ and $\alpha$ provide useful insights into a 
halo's total mass accretion rate, but they cannot distinguish between smooth 
and clumpy accretion. In Fig.~\ref{fig:frequency_of_mergers} we focus 
specifically on a halo's merger history by considering the likelihood that a 
halo of a given $M_{\rm vir}$ (left hand panel) or $z_{\rm form}$ (right hand 
panel) has experienced \emph{at least one} merger with a mass ratio 
$\delta_{\rm max}$ since $z$=0.6. 

Each halo identified at $z$=0 has a unique merger history, which characterises 
not only how its $M_{\rm vir}$ grows as a function of time but also details of 
mergers it has experienced over time. Using this merger history, we construct 
the distribution of mass ratios of mergers $\delta$ experienced by a halo of a 
given $M_{\rm vir}$ or $z_{\rm form}$ between $0 \le z \la 0.6$. We define
$\delta=M_{\rm acc}(z_i,z_f)/M_{\rm vir}(z_f)$ where $M_{\rm acc}(z_i,z_f)$ is the 
mass of the less massive halo prior to its merging with the more massive halo, 
$M_{\rm vir}(z_f)$ is the virial mass of the more massive 
halo once the less massive halo has merged with it, and $z_i>z_f$ and $z_f$
are the redshifts of consecutive simulation snapshots. The maximum value of
$\delta$ for a given halo gives us its $\delta_{\rm max}$ and we use this to 
compute the fraction of haloes of a given $M_{\rm vir}$ or $z_{\rm form}$
that have $\delta_{\rm max}$ in excess of $10\%$ (filled circles), $20\%$ 
(filled squares) and $50\%$ (filled triangles). 

Fig.~\ref{fig:frequency_of_mergers} reveals that mergers with higher mass 
ratios (i.e. minor mergers) are more common than mergers with lower mass 
ratios (i.e. major mergers), independent of $M_{\rm vir}$ and $z_{\rm form}$,
and that more massive (older) haloes tend to experience more mergers than 
their lower mass (younger) counterparts. For example, the likelihood that a 
$10^{12} h^{-1} \rm M_{\sun}$ galaxy-mass halo experiences a merger with 
$\delta_{\rm  max} > 10\%$ is $\sim 35\%$, compared to $25\% (10\%)$ for 
$\delta_{\rm max} > 20\% (50\%)$. In contrast, the likelihood that a 
$10^{14} h^{-1} \rm M_{\sun}$ cluster-mass halo experiences mergers
with $\delta_{\rm max} > 10\% (20\%, 50\%)$ is $\sim 60\% (40\%,20\%)$.
Interestingly, we find that the fraction of haloes that have experienced 
a merger more significant than $\delta_{\rm max}$ increases with $M_{\rm vir}$ 
approximately as $f(\delta_{\rm max}) \propto M_{\rm vir}^{0.11}$. 

These results are broadly in agreement with the findings of 
\citet{fakhouri10a}. Inspection of the leftmost panel of their Figure 7 
shows the mean number of mergers with mass ratios greater than 1:10 and 1:3 
between $z$=0 and $z \sim 0.6$ increases with increasing halo mass, such 
that a $10^{12} (10^{14}) h^{-1} \rm M_{\sun}$ has a likelihood of 
$\sim 40\%$ ($\sim 80\%$) to have experienced a merger with 
$\delta_{\rm  max} > 10\%$, and a likelihood of $\sim 20\%$ ($\sim 40\%$) 
to have experienced a merger with $\delta_{\rm  max} > 33\%$.\\

\begin{figure}
  \includegraphics[width=8cm]{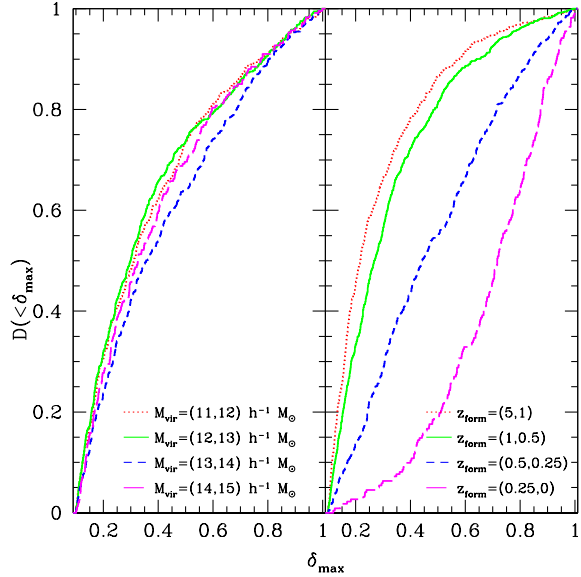}
  \caption{\textbf{Cumulative Distribution of $\delta_{\rm max}$ as a
      Function of Halo Mass and Formation Redshift}. We show how the
    fraction of haloes whose most significant merger's mass ratio
    is less than $\delta_{\rm max}$, as a function of virial mass
    (left hand panel) and formation redshift (right hand panel).
    Note that we select only haloes that have $\delta_{\rm max}
    \geqslant 10\%$, and we consider only mergers between $z$=0.5 and
    $z$=0. In the key, the numbers in brackets correspond to the
    lower and upper bounds in $M_{\rm vir}$ and $z_{\rm form}$.}
  \label{fig:cum_deltamax}
\end{figure}

In Fig.~\ref{fig:cum_deltamax} we show the full (cumulative) distributions of 
$\delta_{\rm max}$ for haloes split into bins according to $M_{\rm vir}$ (left 
hand panel) and $z_{\rm form}$ (right hand panel); note that we consider only 
haloes with $\delta_{\rm max} \geqslant 10\%$. Interestingly this figure reveals
that the probability distribution of $\delta_{\rm max}$ is insensitive to 
$M_{\rm vir}$, but depends strongly on $z_{\rm form}$. For example, the median 
$\delta_{\rm max,med} \simeq 0.3$, independent of $M_{\rm vir}$ whereas it increases
from $\delta_{\rm max,med} \simeq 0.2$ for haloes with $z_{\rm form} \ga 0.5$
to $\delta_{\rm max,med} \simeq 0.4$ for haloes with 
$0.25 \leqslant z_{\rm form} \leqslant 0.5$
and $\delta_{\rm max,med} \simeq 0.7$ for haloes with 
$0 \leqslant z_{\rm form} \leqslant 0.25$.\\

Figs.~\ref{fig:mass_versus_accretion_measures} to
\ref{fig:cum_deltamax} demonstrate that there is a strong correlation
at $z$=0 between a halo's virial mass $M_{\rm vir}$, its formation
redshift $z_{\rm form}$ and the rate at which it has assembled its
mass through accretion and merging over the last $\tau_{\rm dyn}$ or
equivalently $\Delta z \sim 0.6$. We use these results in
\S\ref{sec:linking_mass_assembly_and_dynamical_state}, where we investigate 
the degree to which a halo's $M_{\rm vir}$, $z_{\rm form}$ and mass accretion 
rate affect the degree to which it is in dynamical equilibrium. 

\section{Quantifying Dynamical Equilibrium}
\label{sec:equilibrium}

In this section we describe the two commonly used quantitative measures 
for a halo's dynamical state, the virial ratio $\eta$ and the centre of mass
offset $\Delta r$, and we examine their relationship with virial mass 
$M_{\rm vir}$ and formation redshift $z_{\rm form}$. 

\subsection{The Virial Ratio $\eta$}
\label{ssec:virial_theorem}

The virial ratio $\eta$ is commonly used in cosmological $N$-body simulations 
as a measure of a halo's dynamical state
\citep[e.g.][]{colelacey96,bett07,neto07,Knebe08,Davis11}. It derives from
the virial theorem, 
\begin{equation}
  \label{eq:VirialTheorem}
  \frac{1}{2}\frac{d^2I}{dt^2}=2T+W+E_S,
\end{equation}
\noindent where $I$ is the moment of inertia, $T$ is the kinetic energy,
$W = \Sigma \vec{F}.\vec{r}$ is the virial, and $E_S$ is the surface pressure 
integrated over the bounding surface of the volume within which $I$, $T$ and 
$W$ are evaluated \citep[cf.][]{chandrasekhar61}. Provided the system is
isolated and bounded, the virial $W$ is equivalent to the gravitational
potential energy. While not strictly true for haloes that form in
cosmological $N$-body simulations, the convention has been to evaluate $W$ 
as the gravitational potential energy with this caveat in mind 
\citep[e.g.][]{colelacey96}. We follow this convention and treat $W$ as 
the gravitational potential energy computed using equation (\ref{eq:potential}).

If the system is in a steady state and in the absence of surface pressure,
equation (\ref{eq:VirialTheorem}) reduces to $2\,T+W=0$, which can be written
more compactly as $2\,T/|W|=1$ \citep[e.g.][]{colelacey96}. We refer to the 
ratio $\eta=2\,T/|W|$ as the virial ratio and we expect $\eta \rightarrow 1$ 
for dynamically relaxed haloes. However, we might expect $E_S$ to be important 
for haloes that form in cosmological $N$-body simulations; in this case 
\citet{shaw06} have proposed modifying the virial ratio to obtain
\begin{equation}
  \label{eq:SurfaceVirial}
  \eta'=(2T-E_s)/|W|. 
\end{equation}
\noindent We calculate both $T$ and $W$ using all material within $r_{\rm vir}$, 
while we follow \citet{shaw06} by computing the surface pressure contribution 
from all particles that lie in a spherical shell with inner and outer radii of 
0.8 and 1.0 $r_{\rm vir}$,
\begin{equation}
  \label{eq:pressure}
  P_s = \frac{1}{3V} \Sigma_i(m_iv_i^2);
\end{equation}

\noindent here $V$ corresponds to the volume of this shell and $v_i$ are the 
particle velocities relative to the centre of mass velocity of the halo. 
The energy associated with the surface pressure is therefore $E_s
\simeq 4 \pi r_{med}^3 P_s$ where $r_{\rm med}$ is the median radius of the 
shell.\\

\begin{figure}
  \includegraphics[width=8cm]{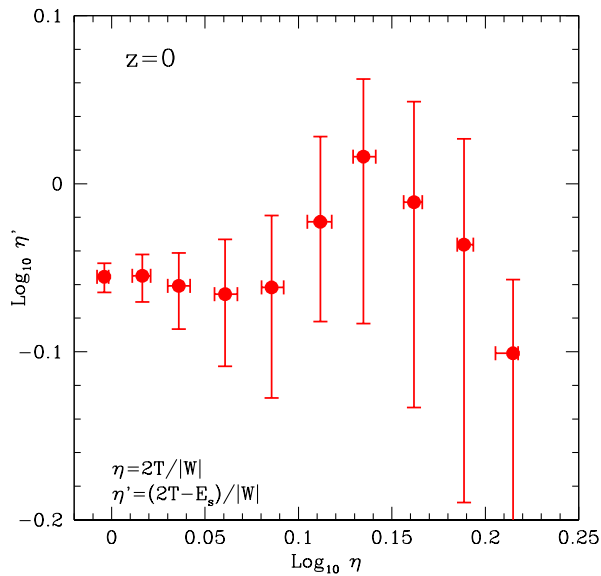}
  \caption{\textbf{Correlation between Virial Ratios $\eta$ and $\eta'$}. 
    We bin all haloes in our sample at $z$=0 according to their $\eta$ and 
    evaluate the median $\eta'$ within each bin. The upper and lower quartiles
    of the distributions in $\eta$ and $\eta'$ are indicated by bars.}
  \label{fig:eta_vs_etadash_z0}
\end{figure}

Figure~\ref{fig:eta_vs_etadash_z0} shows how the median $\eta$ and $\eta'$ for 
the haloes in our sample compare, with bars indicating the upper and lower 
quartiles of the distributions. We might expect that $\eta' \sim 1$ and 
insensitive to variation in $\eta$; however, this figure reveals that the 
relationship between $\eta$ and $\eta'$ is not so straightforward. Haloes 
that we would expect to be dynamically relaxed, with $\eta \sim 1$, have 
values of $\eta'$<0, suggesting that $E_S$ tends to over-correct. Similar 
behaviour has been noted in both \citet{Knebe08} and \citet{Davis11} for 
high redshift haloes ($z \ga 1$). The relation between the median $\eta$ and 
$\eta'$ is flat $\eta\la 1.25$ but 
rises sharply from $\eta'\sim 0.9$ to peak at $\eta' \sim 1.05$ before 
declining sharply for $\eta\ga 1.4$ to a median of $\eta' \sim 0.8$ in the 
last plotted bin. Interestingly, the width of the $\eta'$ distribution 
increases with $\eta$; if $\eta$ tracks recent major merging activity as we 
expect, then this suggests that $\eta'$ -- and consequently the surface 
pressure correction term $E_S$ -- is sensitive to mergers but in a non-trivial 
way.

\subsection{The Centre-of-Mass Offset $\Delta r$}
\label{ssec:centre_of_mass_offset}

Another commonly used measure of a halo's dynamical state is the centre-of-mass
offset $\Delta r$,

\begin{equation}\label{eq:xoff}
  \Delta r = \frac{|\vec{r}_{\rm cen} - \vec{r}_{\rm cm}|}{r_{\rm vir}}, 
\end{equation}

\noindent which measures the separation between a halo's centre-of-density
$\vec{r}_{\rm cen}$ (calculated as described in \S\ref{ssec:halo_properties}) 
and its centre-of-mass, calculated using all material within $r_{\rm vir}$, 
normalised by $r_{\rm vir}$ 
\citep[cf.][]{crone96,thomas98,thomas01,neto07,maccio07,donghia07}. 
$\Delta r$ is used as a substructure statistic, providing an 
estimate of a halo's deviations from smoothness and spherical symmetry. The
expectation is that the smaller the $\Delta r$, the more relaxed the halo; 
for example, \citet{neto07} define dynamically relaxed haloes to be those with
$\Delta r \leq 0.07$, while \citet{donghia07} adopt $\Delta r \leq 0.1$. 
\citet{maccio07} favoured a more conservative $\Delta r \leq 0.04$ based on a 
thorough analysis.\\

\begin{figure}
  \includegraphics[width=8cm]{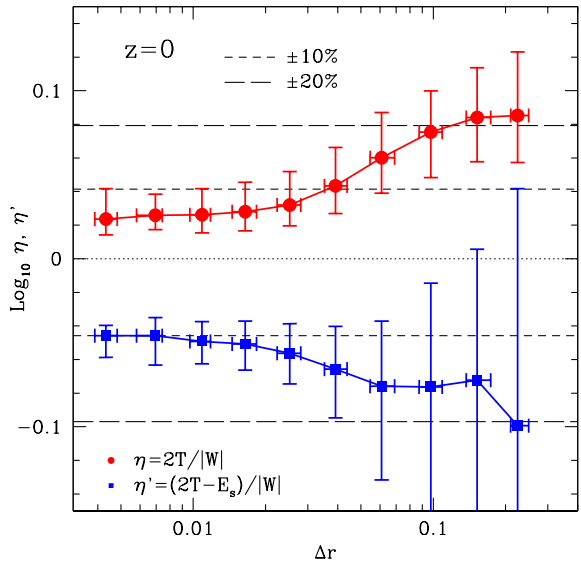}
  \caption{\textbf{Correlation between Centre-of-Mass Offset $\Delta
      r$ and Virial Ratios $\eta$ and $\eta'$}.  We can clearly see the
    relation which is confirmed by measuring a Spearman rank
    coefficient of 0.45 whereas we find an anti-correlation with
    Spearman rank coefficient of -0.18 for $\eta'$.}
  \label{fig:deltar_vs_eta}
\end{figure}

We can get a sense of how well $\Delta r$ measures the dynamical state of 
a halo by comparing it to $\eta$ and $\eta'$. In Figure~\ref{fig:deltar_vs_eta}
we plot the median $\eta$ and $\eta'$ (filled circles and squares respectively)
against the median $\Delta r$; as before, bars indicate the upper 
and lower quartiles of the distributions. This figure shows that both $\eta$ 
and $\eta'$ correlate with $\Delta r$ -- but in different senses; as $\Delta r$
increases, $\eta$ increases while $\eta'$ decreases. The increase (decrease) is
a gradual one; for example, for $\Delta r \la 0.04$, the median $\eta$ is flat
with a value of $\sim 1.05$, but for $\Delta r \ga 0.04$ there is a sharp
increase and $\Delta r \ga 0.1$, $\eta \sim 1.2$. Although direct comparison 
is difficult, a similar trend can be gleaned from Figure 2 of \citet{neto07}. 
We use the Spearman rank correlation coefficient to assess the strength of 
the correlation between $\Delta r$ and $\eta (\eta')$ \citep[cf.][]{Kendall90},
and find strong positive and negative correlations for $\eta$ (Spearman rank 
coeffecient $r$=0.97) and $\eta'$ ($r$=-0.95) respectively.

This is suggestive -- as we show below, $\Delta r$ correlates more strongly 
with merging activity than either of $\eta$ or $\eta'$ (cf. 
Figure~\ref{fig:deltar_vs_mah} in
\S\ref{sec:linking_mass_assembly_and_dynamical_state}). 
Both $\Delta r$ and $\eta$ increase with strength of merging activity, whereas 
$\eta'$ appears to be over-corrected by $E_S$ (as we have noted above). From
this we conclude that $E_S$ (as we evaluate it) correlates with significant
merger activity, which is confirmed by a Spearman rank coefficient of 0.38 for 
the correlation between $E_s$ and $\delta_{\rm max}$.

Interestingly \citet{Davis11} examined the correlation between 
$\Delta r$ and $\eta'$ for high redshift haloes ($z\ga\,6$) and 
noted a tendency for haloes with small values of $\eta'$ to have larger
values of $\Delta r$. Inspection of their Figure 4 shows 
that this is true for haloes with $0.4 \la \Delta r \la 10$; for 
$\Delta r \la 0.4$ the relation with $\eta'$ is flat. \citet{Davis11}
argue that, because there is no systematic shift in $\eta'$ for $\Delta r<0.1$,
$\Delta r$ is not a useful measure of dynamical state at high redshifts.

\subsection{Dependence of Dynamical State on $M_{\rm vir}$ and $z_{\rm form}$}

\begin{figure}
  \includegraphics[width=8cm]{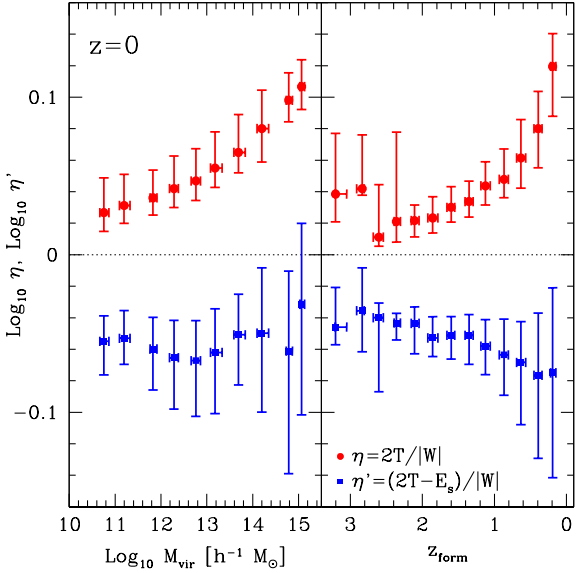}
  \caption{\textbf{Relationship between Dynamical State and Halo Mass and
      Formation Redshift.} For haloes identified at $z$=0, we 
    plot the median $\eta$ and $\eta'$ versus $M_{\rm vir}$ 
    (left hand panel) and $z_{\rm form}$ (right hand panel) using equally 
    spaced bins in $\rm Log_{10} M_{\rm vir}$ and $z_{\rm form}$. Data points and 
    bars correspond to medians and upper and lower quartiles.}
\label{fig:virial_ratio_measures}
\end{figure}

\begin{figure}
  \includegraphics[width=8cm]{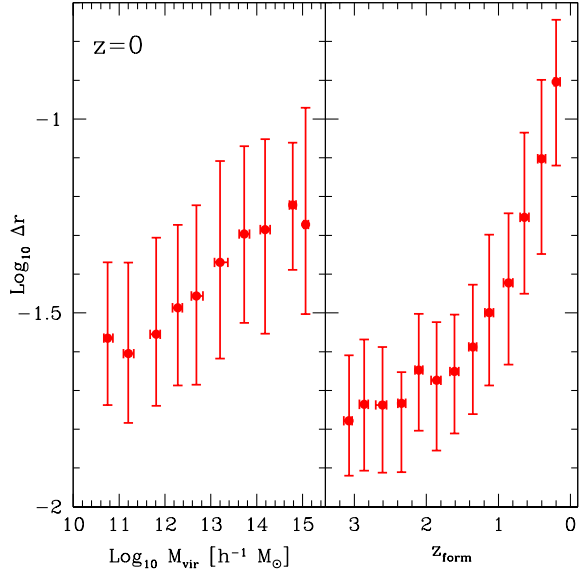}
  \caption{\textbf{Relationship between Centre-of-Mass Offset, Halo Mass and
      Formation Redshift.} For haloes identified at $z$=0, we 
    plot the median centre-of-mass offset $\Delta r$ versus $M_{\rm vir}$ 
    (left hand panel) and $z_{\rm form}$ (right hand panel) using equally 
    spaced bins in $\rm Log_{10} M_{\rm vir}$ and $z_{\rm form}$. Data points and 
    bars correspond to medians and upper and lower quartiles.}
\label{fig:deltar_measures}
\end{figure}

In Figures~\ref{fig:virial_ratio_measures} and \ref{fig:deltar_measures} 
we examine how $\eta$, $\eta'$ and $\Delta r$ vary with $M_{\rm vir}$ (left
hand panels) and $z_{\rm form}$ (right hand panels) for the halo population 
at $z$=0. Haloes are sorted in bins of equal width in mass 
($\Delta \log_{10} M$=0.5 dex) 
and redshift ($\Delta z$=0.25), and we plot the median $\eta$/$\eta'$/$\Delta
r$ within each bin against the median $M_{\rm vir}$/$z_{\rm form}$; bars
indicate the upper and lower quartiles of the respective distributions. For 
reference, we also plot a horizontal dotted line in each panel of 
Figure~\ref{fig:virial_ratio_measures} to indicate a virial ratio of unity. 

\begin{figure}
  \includegraphics[width=8cm]{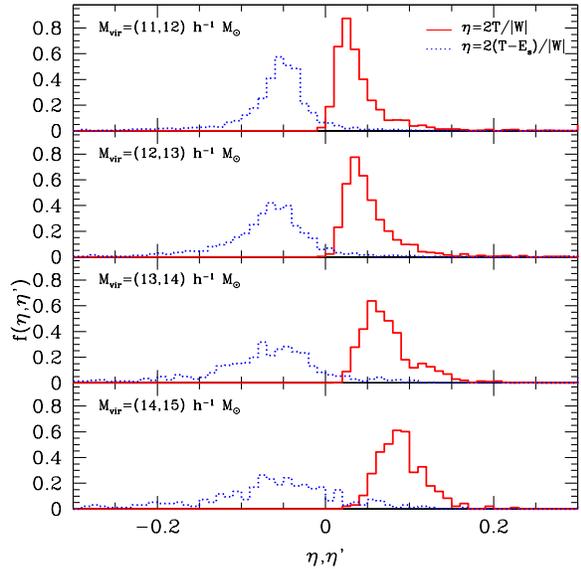}
  \caption{\textbf{Distribution of virial ratios}. 
    Here we show the correlation between halo mass and virial ratio
    $\eta$ and $\eta'$ at redshift $z=0$.}
  \label{fig:eta_vs_mass_z0}
\end{figure}

Because more massive haloes tend to form at later times, and because these 
haloes tend to assemble a larger fraction of their mass more recently, we 
expect that $\eta$ and $\Delta r$ should increase with increasing $M_{\rm vir}$
and decreasing $z_{\rm form}$. This is borne out in Figures~\ref{fig:virial_ratio_measures} and \ref{fig:deltar_measures}. We find that the mean and median 
$\Delta r$ increases steadily with increasing $M_{\rm vir}$ as
\begin{equation}
 \left<\log_{10} \Delta r\right> = -1.47 + 0.08 \log_{10} M_{12}
\end{equation}
\noindent and 
\begin{equation}
 {\mathrm {Med}} \log_{10} \Delta r = -1.49 + 0.09 \log_{10} M_{12}
\end{equation}
\noindent where, as before, $M_{12}$ is $M_{\rm vir}$ in units of 
$10^{12} h^{-1}{\rm M}_{\sun}$. This is consistent with the result of 
\citet[][see their Figure 9]{thomas01}, who found a similar trend for 
$\Delta r$ to increase with $M_{180}$ for a sample of cluster mass haloes 
($10^{13} \la M_{180}/(h^{-1} {\rm M_{\sun}}) \la 10^{15}$) in a $\tau$CDM model. 
Their typical values of $\Delta r$ are offset to higher values than we find, 
but this can be understood as an effect of $\Lambda$, the merging rate being 
suppressed in the $\mathrm\Lambda$CDM model compared to the $\tau$CDM model. 
Similarly, $\Delta r$ varies strongly with $z_{\rm form}$; for $z_{\rm form} 
\ga 1$ we find that $\Delta r \propto (1+z)^{-0.65}$ compared to 
$\Delta r \propto (1+z)^{-0.1}$ for $z_{\rm form} \ga 1$.

The mean and median $\eta$ exhibit similar behaviour, increasing with 
increasing $M_{\rm vir}$, albeit weakly, as,
\begin{equation}
 \left<\log_{10} \eta \right> = 0.05 + 0.016 \log_{10} M_{12}
\end{equation}
\noindent and 
\begin{equation}
 {\mathrm {Med}} \log_{10} \eta = 0.04 + 0.019 \log_{10} M_{12}.
\end{equation}
This means that $\eta$ is systematically greater than unity for all 
$M_{\rm vir}$ that we consider -- $\eta \sim 1.15$ for a typical 
$10^{12} h^{-1} {\rm M_{\sun}}$ halo, compared to $\eta \sim 1.25$ for a 
typical $10^{15} h^{-1} {\rm M_{\sun}}$ halo. The same gradual increase in
$\eta$ with decreasing $z_{\rm form}$ is also apparent.

As we might have anticipated from inspection of Figures
\ref{fig:eta_vs_etadash_z0} and \ref{fig:deltar_vs_eta}, $\eta'$ 
is systematically smaller than unity. Its variation with $M_{\rm vir}$ is
negligible ($\propto M_{\rm vir}^{0.0004}$; a little surprising, when compared to 
$\propto M_{\rm vir}^{0.015}$ at $z$=1, as reported by \citealt{Knebe08}) 
but there is a trend for the median $\eta'$ to decrease with decreasing 
$z_{\rm form}$. This makes sense because $E_S$ increases with the significance 
of recent mergers and haloes that have had recent major mergers tend to have 
smaller $z_{\rm form}$. This effect is also noticeable in the width of the 
$\eta'$ distributions in each bin (as measured by the bars), which are 
larger than than the corresponding widths of the $\eta$ distribution.

We look at this effect in more detail by plotting the distributions of $\eta$
and $\eta'$ shown in Figure~\ref{fig:eta_vs_mass_z0}. Here it is readily 
apparent that there is a systematic shift towards larger $\eta$ as $M_{\rm vir}$
increases. Interestingly the $\eta'$ distribution remains centred on 
$\eta' \sim 0.9$, but it spreads with increasing $M_{\rm vir}$; again, this 
suggests the sensitivity of $\eta'$ to recent merging activity.\\

Figs.~\ref{fig:eta_vs_etadash_z0} to \ref{fig:eta_vs_mass_z0} demonstrate 
that there is a strong correlation at $z$=0 between a halo's virial mass 
$M_{\rm vir}$, its formation redshift $z_{\rm form}$ and its dynamical state, 
as measured by the virial ratio $\eta$ and the centre-of-mass offset $\Delta r$.
In contrast, the correlation with $\eta'$ is more difficult to interpret,
especially when $\eta$ is large. In these cases, we expect significant merging
activity and as we note above, the correction by the surface pressure term
$E_S$ increases the width of the original $\eta$ distribution by a factor of 
$\sim$ 2-3. It's also noteworthy that the median $\eta'$ is systematically 
offset below unity. For this reason we argue that $\eta'$ is not as useful a 
measure of a halo's dynamical state as $\eta$.

\section{Linking Mass Assembly and Dynamical State}
\label{sec:linking_mass_assembly_and_dynamical_state}

We have established quantitative measures of a halo's mass assembly and
merging history and its dynamical state in the previous two sections, and we
have investigated how these relate separately to a halo's virial mass 
$M_{\rm vir}$ and its formation redshift $z_{\rm form}$. In this final section
we examine the relationship between a halo's mass assembly history and its
dynamical state directly.

\begin{figure}
  \includegraphics[width=8cm]{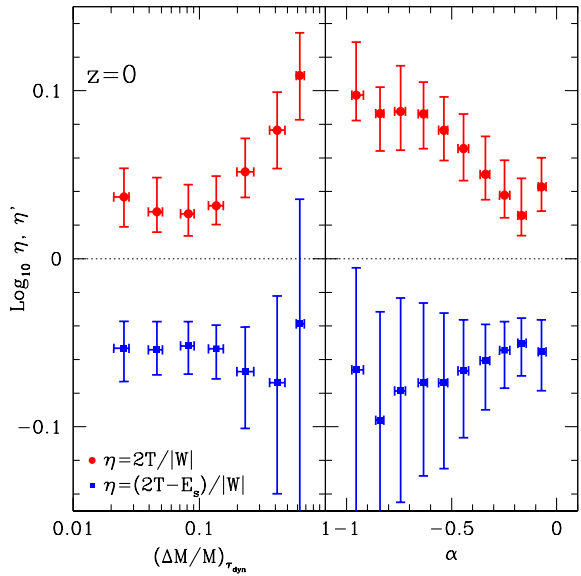}
  \caption{\textbf{Relationship between Recent Mass Accretion and the
      Virial Ratio}. Here we investigate how $(\Delta M/M)_{\tau_{\rm
        dyn}}$, the fraction of mass accreted over $\tau_{\rm dyn}$
    (left hand panel), and $\alpha$, the mean accretion rate over
    $\tau_{\rm dyn}$ (right hand panel), correlate with the standard
    ($\eta$, filled circles) and corrected ($\eta'$, filled squares)
    virial ratio. Data points correspond to medians and bars
    correspond to the upper and lower quartiles.}
  \label{fig:virial_ratio_versus_accretion}
\end{figure}

\begin{figure}
  \includegraphics[width=8cm]{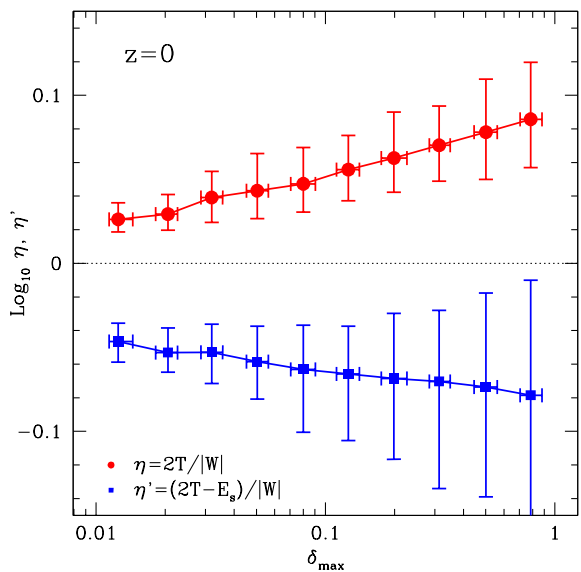}
  \caption{\textbf{Relationship between Most Significant Recent Merger and
      the Virial Ratio}. Here we investigate how $\delta_{\rm max}$,
    which measures the mass ratio of the most significant recent merger 
    since $z$=0.5, correlates with the standard ($\eta$) and corrected 
    ($\eta'$) virial ratios respectively. Filled circles (squares) 
    correspond to medians of $\eta$ ($\eta'$), while bars
    indicate the upper and lower quartiles.}
  \label{fig:virial_ratio_versus_deltamax}
\end{figure}

In Figures~\ref{fig:virial_ratio_versus_accretion} and
\ref{fig:virial_ratio_versus_deltamax} we show explicitly how a halo's
recent mass accretion and merging history impacts on its virial ratio.
As in section~\ref{sec:hierarchical}, we quantify a halo's mass
accretion history by $(\Delta M/M)_{\tau_{\rm dyn}}$, the fractional
increase in a halo's mass over the period $\tau_{\rm dyn}$ (equivalent
to a redshift interval $\Delta z \simeq 0.6$ at $z$=0), and
$\alpha$, the mean accretion rate over the period $\tau_{\rm
  dyn}$). We use $\delta_{\rm max}$, the mass ratio of the most
significant merger experienced by the halo over $\tau_{\rm dyn}$, to
characterise a halo's recent merging history.

\begin{figure}
  \includegraphics[width=8cm]{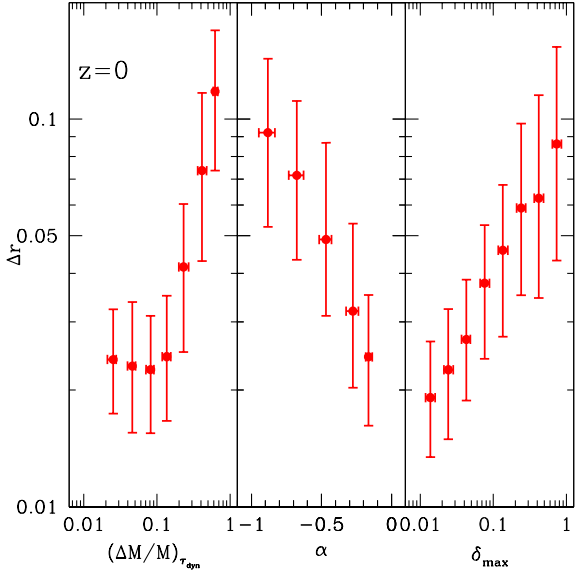}
  \caption{\textbf{Correlation between centre-of-mass offset 
      $\Delta r$ and Recent Merging and Accretion History}. 
    Here we examine whether $\Delta r$ correlates with the fraction of
    mass accreted over $\tau_{\rm dyn}$, $(\Delta M/M)_{\tau_{\rm
	dyn}}$, the mean accretion rate $\alpha$ and the most
    significant merger $\delta_{\rm max}$. Filled circles correspond 
    to medians in the respective bins; bars represent the upper and lower
    quartiles of the distribution.}
  \label{fig:deltar_vs_mah}
\end{figure}

We expect that the standard virial ratio $\eta$ should increase with
increasing mass accretion rate and decreasing mass ratio of most
significant merger, which is in good agreement with the behaviour that
we observe. In particular, the median variation of $\eta$ with
$\alpha$ and $\delta_{\rm max}$ can be well approximated by
$\log_{10} \eta \simeq 0.004 - 0.126 \alpha$ and 
$\eta \simeq 1.2\delta_{\rm max}^{1.1}$; the corresponding 
variation of $\Delta r$ can be well approximated by
$0.01-0.1\alpha$ and $0.1\delta_{\rm max}^{0.3}$.

Interestingly we note that the median corrected virial
ratio $\eta'$ declines with increasing mass accretion rate and mass
ratio of most significant merger. Both correlations indicate
that merger events lead to a state that is less virialised, but, as we have
noted already, the inclusion of the surface pressure term over-corrects the 
virial ratio. We see in Figure~\ref{fig:virial_ratio_versus_accretion} that 
for $(\Delta M/M)_{\tau_{\rm dyn}} \la 0.2$, both the median $\eta$ and 
$\eta'$ are flat; $\eta \sim 1.05$ whereas the median $\eta' \sim 0.85$. 
Above $(\Delta M/M)_{\tau_{\rm dyn}} \sim 0.2$, the median $\eta$ increases 
sharply whereas it is the width of the $\eta'$ distribution that shows the 
sharp increase. Comparison with Figure~\ref{fig:virial_ratio_versus_deltamax} 
provides further insight -- the median $\eta$ ($\eta'$) shows a gradual 
increase (decrease) with increasing $\delta_{\rm max}$, starting at $\eta \sim
1.05$ ($\eta' \sim 0.9$) for $\delta_{\rm max} \sim 0.02$. For $\delta_{\rm
  max}$=0.1, $\eta \sim 1.1$ ($\eta' \sim 0.85$). However, whereas the width
of the $\eta$ distribution is largely insensitive to $\delta_{\rm max}$, the
width of the $\eta'$ distribution increases rapidly, bearing out our
observations in the previous section.

In Figure~\ref{fig:deltar_vs_mah} we show how $\Delta r$ varies with
$(\Delta M/M)_{\tau_{\rm dyn}}$, $\alpha$ and $\delta_{\rm max}$. This reveals
that $\Delta r$ increases with increasing mass accretion rate and mass ratio 
of most significant recent merger, as we would expect. Although the scatter 
in the distribution is large, we can identify the remnants of recent major
mergers ($\delta_{\rm max} \ga 30\%$) as haloes with $\Delta r
\ga 0.06$. Haloes that have had relatively quiescent recent mass
accretion histories ($(\Delta M/M)_{\tau_{\rm dyn}} \la 0.1$,
$\delta_{\rm max} \la 10\%$) have $\Delta r \la 0.04$.

\paragraph*{Merging Timescale \& Dynamical State}

We conclude our analysis by investigating the timescale over which the effect
of a merger can be observed in the virial ratio $\eta$ and the centre-of-mass
offset $\Delta r$.

\begin{figure}
  \includegraphics[width=8cm]{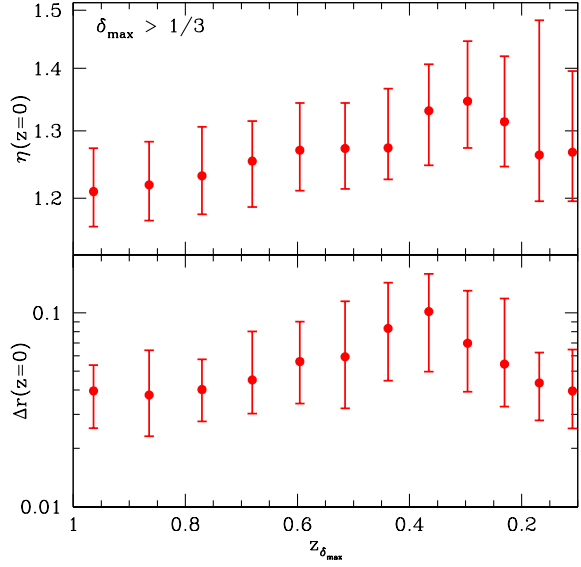}
  \caption{\textbf{Relationship between $\eta$, $\Delta r$ and 
      $z_{\delta_{\rm max}}$, the redshift of the most significant recent major 
      merger}. We identify all haloes in our sample at $z$=0 with 
    $\delta_{\rm max} \gtrsim 1/3$ and identify the redshift $z_{\delta_{\rm max}}$ 
    at which $\delta_{\rm max}$ occurred. Both $\eta$ and $\Delta r$ are 
    evaluated at z=0. Filled circles and bars correspond to medians and upper 
    and lower quartiles.}
  \label{fig:time_dep_dynstate}
\end{figure}

In Figure~\ref{fig:time_dep_dynstate} we investigate how a typical halo's
$\eta$ (upper panel) and $\Delta r$ (lower panel), measured at $z=0$, 
correlates with the redshift at which the halo suffered it's most significant 
merger, $z_{\delta_{\rm max}}$. For clarity, we focus on haloes for which 
$\delta_{\rm max}>1/3$, although we have verified that our results are not 
sensitive to the precise value of $\delta_{\rm max}$ that we adopt; filled 
circles correspond to medians and bars indicate upper and lower quartiles. The 
median $\eta$ increases with decreasing $z_{\delta_{\rm max}}$ for 
$z_{\delta_{\rm max}} \gtrsim 1$ before peaking at $z_{\delta_{\rm max}} \simeq 0.3$
and declining at lower $z_{\delta_{\rm max}}$. The median $\Delta r$ shows a similar
steady increase with decreasing $z_{\delta_{\rm max}}$ below $z_{\delta_{\rm max}} \sim 
0.8$ although there is evidence that it peaks at $z_{\delta_{\rm max}} \simeq 0.4$ 
before declining at lower $z_{\delta_{\rm max}}$. The redshift interval 
corresponding to $z_{\delta_{\rm max}} \simeq 0.3$ represents a time interval of 
$\Delta t \simeq 4.3\,{\rm Gyrs}$ or $\sim 1.5\,\tau_{\rm dyn}$. 

This is consistent with the finding of \citet{tormen97}, who examined the 
velocity dispersion $v_{\rm rms}$ of material within $r_{\rm vir}$ of simulated 
galaxy cluster haloes (see their Figure 5). They noted that merging leads to 
an increase in $v_{\rm rms}$ of the main (host) halo because the merging 
sub-halo acquires kinetic energy as it falls in the potential well of the more 
massive main halo. The peak in $v_{\rm rms}$ corresponds to the first 
pericentric passage of the subhalo, after which $v_{\rm rms}$ declines because 
subsequent passages are damped, and so the main halo relaxes. This will occur
on a timescale of order $\sim 1-2\,\tau_{\rm dyn}$, which is consistent with the
peak in $\eta$ at $z_{\delta_{\rm max}} \simeq 0.3$. We would expect to see a peak
in $\Delta r$ on roughly the merging timescale $\tau_{\rm merge}$, which as we 
noted in \S~\ref{sec:hierarchical} is comparable to $1-2\,\tau_{\rm dyn}$
\citep[cf.][]{boylanetal08}.\\

\begin{figure}
  \includegraphics[width=8cm]{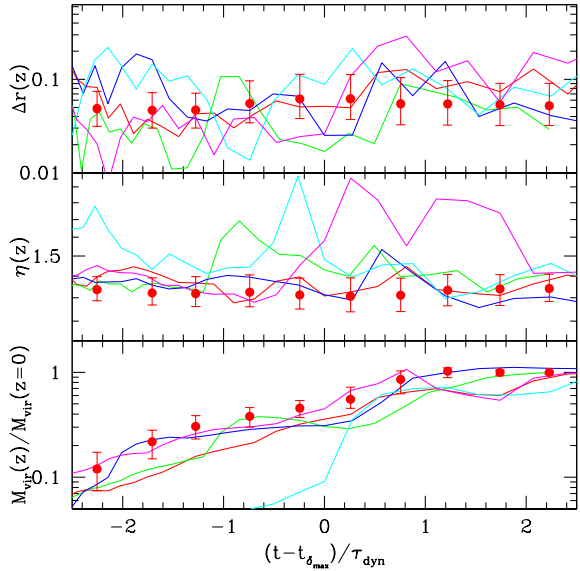}
  \caption{\textbf{Response of $M_{\rm vir}$, $\eta$ and $\Delta r$ to a major 
      merger.} We include all haloes with $\delta_{\rm max} \gtrsim 1/3$ at 
    $z$=0 and plot the redshift variation of $M_{\rm vir}$, $\eta$ and 
    $\Delta r$ against the time since the major merger, normalised by the
    dynamical time of the halo at the redshift at which the merger occurred.
    Filled circles and bars correspond to medians and upper and lower quartiles,
    while curves correspond to the histories of 5 individual haloes.}
  \label{fig:dynstate_merge_tree}
\end{figure}

We can take this analysis a little further by looking at the detailed evolution
of $\eta$ and $\Delta r$ over time. In Figure~\ref{fig:dynstate_merge_tree} we 
plot the redshift variation of 
$M_{\rm vir}$ (normalised to its value at $z$=0; lower panel), $\eta$ 
(middle panel) and $\Delta r$ (upper panel) against the time since major 
merger, normalised by the dynamical time $\tau_{\rm dyn}$ estimated
at the redshift at which the merger occurred, $z_{\delta_{\rm max}}$. Medians and
upper and lower quartiles are indicated by filled circles and bars. For 
illustrative purposes, we show also the redshift variation of $M_{\rm vir}$, 
$\eta$ and $\Delta r$ for a small subset of our halo sample (red, blue, green,
cyan and magenta curves). As in Figure~\ref{fig:time_dep_dynstate}, we adopt 
$\delta_{\rm max}\gtrsim 1/3$.

Our naive expectation is that both $\eta$ and $\Delta r$ should increase in
response to the merger, peak after $\Delta\,t \simeq \tau_{\rm dyn}$ and then
return to their pre-merger values. If this behaviour is typical, then we 
expect pronounced peaks in the median values of $\eta$ and $\Delta r$ at 
$\Delta\,t/\tau_{\rm dyn} \simeq 1$. However, it is evident from 
Figure~\ref{fig:dynstate_merge_tree} that there is no significant 
difference between the median $\eta$ and $\Delta r$ pre- and post-major merger, 
and so our naive expectation is not borne out by our results.

This is not surprising if one inspects histories for $\eta$ and $\Delta r$ for 
individual haloes, in the spirit of \citet{tormen97}; $\eta$ and $\Delta r$ 
increase following a major merger, but the behaviour is noisy (reflecting e.g. 
differences in orbital parameters of merging subhaloes, the redshift dependent 
virial radius, dependence on environment, etc...) and the timescale of the 
response varies from halo to halo -- simply averaging or taking the median 
washes any signal away. Nevertheless it is worth looking at this in more detail,
which we shall do in a forthcoming paper.

\section{Summary}
\label{sec:summary}

The aim of this paper has been to quantify the impact of a dark matter halo's 
mass accretion and merging history on two measures of dynamical state
that are commonly used in cosmological $N$-body simulations, namely the virial 
ratio $\eta=2T/|W|$ \citep[cf.][]{colelacey96} and the centre-of-mass 
offset $\Delta r=|\vec{r}_{\rm cen}-\vec{r}_{cm}|/r_{\rm vir}$ 
\citep[cf.][]{crone96,thomas98,thomas01}. The virial ratio $\eta$ derives from
the virial theorem and and the expectation is that $\eta \rightarrow 1$ for
dynamically equilibrated haloes. The centre-of-mass offset $\Delta r$ can be
regarded as a substructure statistic \citep{thomas01} that provides a convenient
measure of how strongly a halo deviates from smoothness and spherical symmetry.
We expect a halo's dynamical state and its mass assembly history to correlate
closely. Understanding how is important because the degree to which a halo is 
dynamically equilibrated affects the reliability with which we can measure 
characteristic properties of its structure, such as the concentration parameter 
$c_{\rm vir}$ \citep[e.g.][]{maccio07,neto07,Prada11}, and kinematics, such 
as the spin parameter $\lambda$ \citep[e.g.][]{gardner01,donghia07,Knebe08}. 
For this reason, it is desirable to establish quantitatively how well they 
correlate and to assess how $\eta$ and $\Delta r$ can help us to characterise
how quiescent or violent a halo's recent mass assembly history has been.

Our key results are that $\eta$ and $\Delta r$ show strong positive
correlations with each other (cf. Figure~\ref{fig:deltar_vs_eta}) -- as $\eta$
increases for a halo, so too does $\Delta r$ -- and that both are useful
indicators of a halo's mass recent mass accretion and merging history. For
example, $\eta$ and $\Delta r$ correlate strongly with $\delta_{\rm max}$,
which measures the significance of a halo's recent merging activity; haloes 
with $\eta \la 1.1$ (cf. Figure~\ref{fig:virial_ratio_versus_deltamax}) 
and $\Delta r \la 0.04$ (cf. Figure~\ref{fig:deltar_vs_mah}) will have
quiescent recent mass assembly histories -- they are unlikely to have suffered 
mergers with mass ratios greater than 1:10 over the last few dynamical times.

In contrast, interpreting the corrected virial ratio $\eta'$=$(2T-E_S)/|W|$, where
$E_s$ is the surface pressure energy, is problematic (at least insofar as we 
have implemented it here, which follows the prescription of \citealt{shaw06} 
and has been applied in \citealt{Knebe08} and \citealt{Davis11}). In principle, 
$\eta'$ should account for the approximation that is made when we define a 
halo to be a spherical overdensity of $\Delta_{\rm vir}$ times the critical 
density at a particular redshift. As we noted in \S~\ref{sec:sims}, haloes
are more complex structures than this simple working definition gives them 
credit for, and by defining the halo's extent by the virial radius $r_{\rm vir}$ the
likelihood is that material that belongs to the halo will be neglected. By correcting 
the virial ratio $\eta$ for what is effectively a truncation of the true halo, the 
corrected virial ratio $\eta'$ takes account of the ``missing'' kinetic energy. However, 
our results imply that the correction itself (the surface pressure energy $E_S$) is 
sensitive to a halo's merging history, and that it increases with increasing 
$\delta_{\rm max}$ (cf. Figure~\ref{fig:virial_ratio_versus_deltamax}). For this reason 
we would caution against the use of $\eta'$ to identify dynamically relaxed haloes, at
least in the form that is currently used.

Interestingly, we find that systems with violent recent mass assembly 
histories (most significant merger with a mass ratio $\delta_{\rm max} \ga 1/3$
between $0 \la z \la 1$) have values of $\eta$ and $\Delta r$ (as measured at $z$=0) 
that peak at $z_{\delta_{\rm max}} \simeq 0.3-0.4$, which corresponds 
to a timescale of $\sim 1.5\,\tau_{\rm dyn}$ (cf. Figure~\ref{fig:time_dep_dynstate}).
This is consistent with the earlier analysis of \citet{tormen97}, who found that
the velocity dispersion $v_{\rm rms}$ of material within the virial radius -- which is
linked to the virial ratio $\eta$ -- peaks on first closest approach of the merging 
sub-halo with the centre of the more massive host halo. This should occur on a timescale
of $\sim 1-2\,\tau_{\rm dyn}$, after which $v_{\rm rms}$ and $\eta$ should dampen away. 
Similar arguments can be made for $\Delta r$. We note that these arguments can be made
in a statistical sense, but if we look at the merging histories of individual haloes, the
behaviour of $\eta$ and $\Delta r$ is much more complex, and as we demonstrate
a simple timescale for their response to a major merger is difficult to define (cf. 
Figure~\ref{fig:dynstate_merge_tree}). We shall return to this topic in future work.\\

What is the significance of these results? Structure formation proceeds 
hierarchically in the CDM model and so we expect to find correlations between 
virial mass $M_{\rm vir}$ and formation redshift $z_{\rm form}$ (cf. 
Figure~\ref{fig:mass_versus_zform}), which in turn result in positive 
correlations between $M_{\rm vir}$/$z_{\rm form}$ and $\eta$/$\Delta r$.
(cf. Figures~\ref{fig:virial_ratio_measures} and~\ref{fig:deltar_measures}). 
This means that more massive haloes and those that formed more recently are
also those that are least dynamically equilibrated, a fact that we should be
mindful of when characterising the halo mass dependence of halo properties 
that are sensitive to dynamical state (e.g. $c_{\rm vir}$ and $\lambda$). 
It's worth noting that the correlation between $M_{\rm vir}$ and $\eta$ is 
stronger than the correlation between $M_{\rm vir}$ and $\Delta r$; the median 
$\eta$ rises sharply with $M_{\rm vir}$ and there is no overlap between the
width of the distributions of $\eta$ is the lowest and highest mass bins. 
In contrast, the median $\Delta r$ in the highest mass bin lies in the 
high-$\Delta r$ tail of the lowest mass bin. 

This is interesting because $\eta$ as it is usually calculated depends on $W$, 
which is sensitive to the precise boundary of the halo. Correcting for the 
surface pressure term does not appear to help, as we point out -- indeed, the 
surface pressure term itself correlates with merging activity. This points
towards an ambiguity in the use of $\eta$ -- as we note, it rarely if ever 
satisfies $\eta$=1. We discuss this point in a forthcoming paper, but we note 
that even in ideal situations, what one computes for $\eta$ depends on 
$r_{\rm vir}$ \citep[cf.][]{colelacey96,lm01} -- and so applying a flat cut based
on a threshold in $\eta$ alone risks omitting massive haloes that might 
otherwise be considered dynamically equilibrated. For this reason we advocate 
the use of $\Delta r$ in cosmological $N$-body simulations as a more robust 
measure of a halo's dynamical state; its calculation is computationally 
inexpensive, it is well defined as a quantity to measure, and
its interpretation is both clear and straightforward. We find that $\Delta r
\la 0.04$, which corresponds to a $\delta_{\rm max} \la 0.1$, should
be sufficient to pick out the most dynamically relaxed haloes in a simulation
volume at $z$=0.\\

Although our focus has  been fixed firmly on haloes in cosmological 
simulations, we note that our results have observational implications.
Whether or not an observed system -- for example, a galaxy cluster -- is 
in dynamical equilibrium will affect estimates of its dynamical mass if we
assume a luminous tracer population that is in dynamical equilibrium 
\citep[e.g.][]{piffaretti08}. Similarly, studies that seek to reconstruct 
a galaxy cluster's recent merging history tend to use signatures of 
disequilibrium \citep[e.g.][]{cassano10}. The most obvious measure of 
disequilibrium is the centre of mass offset $\Delta r$, or its projected
variant. Although a more careful study in which we mock observe our haloes
(and a seeded galaxy population) is needed, our results suggest that $\Delta r$
could be used to infer the redshift of the last major merger (cf. 
Figures~\ref{fig:deltar_vs_mah} and Figure~\ref{fig:deltar_measures}, although
care must be taken as Figure~\ref{fig:dynstate_merge_tree} reveals). 
Observationally, this would require measurement of, for example, projected 
displacements between gas and dark matter from gravitational lensing and 
X-ray studies. We note that \citet{poole06} have already tested this idea 
using idealised hydrodynamical simulations of mergers between galaxy clusters 
and found that the centroid offset between X-ray and projected mass maps 
captures the dynamical state of galaxy clusters well, but it is interesting 
to extend this idea using cosmological hydrodynamical simulations of galaxy 
groups and clusters. This will form the basis of future work.

\section*{Acknowledgments}

CP acknowledges the support of the STFC theoretical astrophysics 
rolling grant at the University of Leicester. AK is supported by the 
{\it Spanish Ministerio de Ciencia e Innovaci\'on} (MICINN) in 
Spain through the Ramon y Cajal programme as well as the grants AYA 
2009-13875-C03-02, AYA2009-12792-C03-03, CSD2009-00064, and 
CAM S2009/ESP-1496. He further thanks the Aluminum Group for chocolates. 
SRK acknowledges financial support from Swinburne University of Technology's 
Centre for Astrophysics and Supercomputing's visitor programme. He 
acknowledges support by the MICINN under the Consolider-Ingenio, SyeC 
project CSD-2007-00050. The simulations presented in this paper were 
carried out on the Swinburne Supercomputer at the Centre for 
Astrophysics \& Supercomputing, the Sanssouci cluster at the AIP and the
ALICE supercomputer at the University of Leicester.

\vspace{1cm} \bsp

\bibliographystyle{mn2e}

\label{lastpage}

\end{document}